\documentclass[printer]{aa}
\usepackage{graphics,epsfig,lscape,txfonts}
\usepackage{natbib}
\bibpunct{(}{)}{;}{a}{}{,}

\def\cm-2{cm$^{-2}$}

\def\HII{\hbox{H\,{\sc ii}}}

\def\ein{{\it Einstein}}
\def\chandra{{\it Chandra}}

\def\xmm{{XMM-Newton}}
\def\n253{\object{NGC~253}}
\def\m33{\object{M~33}}
\def\mx7{\object{M~33~X$-$7}}
\def\x7{\hbox{X$-$7}}

\newcommand{\ergcm}[1]{$\times10^{#1}$ \hbox{erg cm$^{-2}$ s$^{-1}$}}

\newcommand{\ergs}[1]{$\times10^{#1}$ \hbox{erg s$^{-1}$}}
\newcommand{\oergs}[1]{$10^{#1}$ erg s$^{-1}$}
\newcommand{\hcm}[1]{$\times10^{#1}$ cm$^{-2}$}

\newcommand{\expo}[1]{$\times10^{#1}$}

\newcommand{\nh}{\hbox{$N_{\rm H}$}}

\begin{document}

   \title{\xmm\ survey of the Local Group galaxy \m33
   \thanks{XMM-Newton is an ESA Science Mission 
    with instruments and contributions directly funded by ESA Member
    States and the USA (NASA).}}

   \author{W.~Pietsch\inst{1} \and 
	   Z.~Misanovic\inst{1} \and
	   F.~Haberl\inst{1} \and
	   D.~Hatzidimitriou\inst{2} \and
	   M.~Ehle\inst{3} \and
	   G.~Trinchieri\inst{4}
          }
\institute{Max-Planck-Institut f\"ur extraterrestrische Physik, Giessenbachstra\ss e, 85741 Garching, Germany 
	  \and Department of Physics, University of Crete, P/O. Box 2208, 71003 Heraklion, Crete, Greece
	  \and \xmm\ Science Operations Centre, ESA, Villafranca del Castillo,
	  P.O. Box 50727, 28080 Madrid, Spain
	  \and Osservatorio Astronomico di Brera, via Brera 28, 20121 Milano, Italy  }
     
     \offprints{W.~Pietsch, e-mail: {\tt wnp@mpe.mpg.de}}

   \date{Received; accepted }
   \titlerunning{\xmm\ survey of the Local Group galaxy M 33}

	\abstract{ In an \xmm\ raster observation of the bright Local Group spiral galaxy
\m33\ we study the population of X-ray sources (X-ray binaries, supernova
remnants) down to a 0.2--4.5 keV luminosity of \oergs{35} --
more than a factor of 10 deeper than earlier ROSAT observations. EPIC 
hardness ratios and optical and radio information are used to distinguish between 
different source classes. The survey detects 408 sources in an area of 0.80 square degree.
We correlate these newly detected sources with earlier \m33\ X-ray catalogues and information 
from optical, infra-red and radio wavelengths.
As \m33\ sources we detect 21 supernova remnants (SNR) and 23 SNR candidates, 
5 super-soft sources, and 2 X-ray binaries
(XRBs). There are 267 sources classified as hard, which may either be 
XRBs or Crab-like SNRs in \m33\ or background AGN. The 44 confirmed and candidate SNRs more than 
double the number of X-ray detected SNRs in \m33. 16 of these are proposed as SNR candidates 
from the X-ray data for the first time.
On the other hand, there are several sources not connected to \m33:
five foreground stars, 30 foreground star candidates, 12 active galactic
nucleus candidates,  one background galaxy and one background galaxy
candidate. 
Extrapolating from deep field observations we would
expect 175 to 210 background sources in this field. This indicates that about 
half of the sources detected are sources within \m33.
	
} 
\maketitle

\section{Introduction}
\begin{table*}
\begin{center}
\caption[]{\xmm\ \m33\ observation log (proposal numbers 010264 and 014198).}
\begin{tabular}{rllrrlrlrlr}
\hline\noalign{\smallskip}
\hline\noalign{\smallskip}
\multicolumn{1}{c}{Field} & \multicolumn{1}{c}{Obs. id.} &\multicolumn{1}{c}{Obs. dates} &
\multicolumn{2}{c}{Pointing direction} & \multicolumn{2}{c}{EPIC PN} & 
\multicolumn{2}{c}{EPIC MOS1} & \multicolumn{2}{c}{EPIC MOS2}  \\ 
\noalign{\smallskip}
M33 & & & \multicolumn{2}{c}{RA/DEC (J2000)} 
& \multicolumn{1}{c}{Mode$^{+}$/Filter}  & \multicolumn{1}{c}{T$_{exp}^{\dagger}$}
& \multicolumn{1}{c}{Mode$^{+}$/Filter}  & \multicolumn{1}{c}{T$_{exp}^{\dagger}$}
& \multicolumn{1}{c}{Mode$^{+}$/Filter}  & \multicolumn{1}{c}{T$_{exp}^{\dagger}$}\\
\noalign{\smallskip}
\multicolumn{1}{c}{(1)} & \multicolumn{1}{c}{(2)} & \multicolumn{1}{c}{(3)} & 
\multicolumn{1}{c}{(4)} & \multicolumn{1}{c}{(5)} & \multicolumn{1}{c}{(6)} & 
\multicolumn{1}{c}{(7)} & \multicolumn{1}{c}{(8)} & \multicolumn{1}{c}{(9)} & 
\multicolumn{1}{c}{(10)} & \multicolumn{1}{c}{(11)} \\
\noalign{\smallskip}\hline\noalign{\smallskip}
1  & 0102640101 & 2000-08-04 & 1:33:51.0 & 30:39:37 & FF/medium &  6.29 & PW2/medium &  6.23 & PW2/medium &  6.22 \\
1  & 0141980501 & 2003-01-22 & 1:33:51.0 & 30:39:37 & FF/medium &  1.23 & FF/medium  &  1.51 & FF/medium  &  1.51 \\
1  & 0141980801 & 2003-02-12 & 1:33:51.0 & 30:39:37 & FF/medium &  7.25 & FF/medium  &  8.86 & FF/medium  &  8.87 \\
2  & 0102640201 & 2000-08-04 & 1:34:40.0 & 30:57:48 & EFF/medium & 11.40 & FF/medium  & 15.24 & FF/medium  & 15.24 \\
3  & 0102640301 & 2000-08-07 & 1:33:32.0 & 30:52:13 & EFF/medium &  3.80 & FF/medium  &  4.58 & FF/thin    &  4.58 \\
3  & 0141980401 & 2003-01-24 & 1:33:32.0 & 30:52:13 & FF/medium &  0.00 & FF/medium  &  0.00 & FF/medium  &  0.00 \\
4  & 0102640401 & 2000-08-02 & 1:32:51.0 & 30:36:49 & EFF/thick &  8.66 & FF/thick   & 12.99 & FF/thick   & 12.99 \\
4  & 0141980601 & 2003-01-23 & 1:32:51.0 & 30:36:49 & FF/medium & 11.18 & FF/medium  & 12.56 & FF/medium  & 12.56 \\
5  & 0102640501 & 2001-07-05 & 1:33:02.0 & 30:21:24 & FF/medium &  9.26 & FF/medium  & 11.76 & FF/medium  & 11.76 \\
6  & 0102640601 & 2001-07-05 & 1:34:08.0 & 30:46:06 & FF/medium &  2.86 & FF/medium  &  3.42 & FF/medium  &  3.42 \\
6  & 0141980301 & 2003-07-25 & 1:34:08.0 & 30:46:06 & FF/medium &  4.12 & FF/medium  &  4.82 & FF/medium  &  4.82 \\
7  & 0102640701 & 2001-07-05 & 1:34:10.0 & 30:27:00 & FF/medium &  6.97 & FF/medium  & 10.96 & FF/medium  & 10.95 \\
8  & 0102640801 & 2001-07-07 & 1:34:51.0 & 30:42:22 & FF/medium &     - & FF/medium  &  1.19 & FF/medium  &  1.19 \\
8  & 0102642001 & 2001-08-15 & 1:34:51.0 & 30:42:22 & FF/medium &  8.65 & FF/medium  & 11.15 & FF/medium  & 11.15 \\
9  & 0102640901 & 2001-07-07 & 1:34:04.0 & 30:57:25 & FF/medium &  2.42 & FF/medium  &  3.15 & FF/medium  &  3.15 \\
9  & 0141980201 & 2003-07-11 & 1:34:04.0 & 30:57:25 & FF/medium &  3.74 & FF/medium  &  4.02 & FF/medium  &  4.02 \\
10 & 0102641001 & 2001-07-08 & 1:33:07.0 & 30:45:02 & FF/medium &  1.25 & FF/medium  &  8.33 & FF/medium  &  8.37 \\
10 & 0141980101 & 2003-07-11 & 1:33:07.0 & 30:45:02 & FF/medium &  5.45 & FF/medium  &  5.75 & FF/medium  &  5.77 \\
11 & 0102641101 & 2001-07-08 & 1:32:46.0 & 30:28:19 & FF/medium &  8.10 & FF/medium  & 10.36 & FF/medium  & 10.36 \\
12 & 0102641201 & 2000-08-02 & 1:33:38.0 & 30:21:49 & EFF/thick & 11.34 & FF/thick   &  3.67 & FF/thick   &  3.66 \\
12 & 0141980701 & 2003-01-24 & 1:33:38.0 & 30:21:49 & FF/medium &  4.40 & FF/medium  &  5.72 & FF/medium  &  5.73 \\
13 & 0102642101 & 2002-01-25 & 1:34:34.0 & 30:34:11 & FF/medium & 10.00 & FF/medium  & 12.27 & FF/medium  & 12.27 \\
14 & 0102642201 & 2002-01-25 & 1:34:56.0 & 30:50:52 & FF/medium & 11.60 & FF/medium  & 13.87 & FF/medium  & 13.87 \\
15 & 0102642301 & 2002-01-27 & 1:33:33.0 & 30:33:07 & FF/medium & 10.00 & FF/medium  & 12.27 & FF/medium  & 12.27 \\
\noalign{\smallskip}
\hline
\noalign{\smallskip}
\end{tabular}
\label{observations}
\end{center}
Notes:\\
$^{ +~}$: FF: full frame, EFF: extended full frame, PW2: partial window 2 \\
$^{ {\dagger}~}$: Exposure time in units of ks after screening for high background (see text)\\
\end{table*}
The Local Group Sc galaxy \m33\ is located at a distance of 795 kpc
\citep[][ i.e. 1\arcsec\ corresponds to 3.9 pc and the flux to luminosity conversion factor
is 7.6\expo{49} cm$^2$]{1991PASP..103..609V}, is seen under a
relatively low inclination of 56$^{\circ}$ \citep{1989AJ.....97...97Z}
and the optical extent can be approximated by an inclination-corrected
D$_{25}$ ellipse with large diameter 
of 64\farcm4 and axes ratio of 1.66
\citep{1991trcb.book.....D,1988ngc..book.....T}.
With its moderate Galactic foreground absorption 
\citep[\nh $= 6\times10^{20}$\,cm$^{-2}$, ][]{1992ApJS...79...77S} \m33\ is
ideally suited to study the X-ray source population and diffuse emission in a
nearby spiral galaxy. The \ein\ X-ray Observatory detected diffuse emission from hot
gas in \m33\ and 17 unresolved sources 
\citep{1981ApJ...246L..61L,1983ApJ...275..571M,1988ApJ...329.1037T}. 
First ROSAT HRI and PSPC observations
revealed 57 sources and confirmed the detection of diffuse X-ray emission which
may trace the spiral arms within a 10' radius around the nucleus 
\citep{1995ApJ...441..568S,1996ApJ...466..750L}.
Combining all archival ROSAT observations of the field,
\defcitealias{2001A&A...373..438H}{HP01} 
\citet[][ hereafter \citetalias{2001A&A...373..438H}]{2001A&A...373..438H}
found 184 X-ray sources within a 50' radius around the nucleus,
identified some of the sources by correlations with previous X-ray, optical and
radio catalogues, and in addition classified sources according to their X-ray
properties. They found candidates for super-soft X-ray sources (SSS), 
X-ray binaries (XRBs), supernova remnants (SNRs), foreground stars and active
galactic nuclei (AGN). 

To follow up on these findings, we proposed as part of the telescope
scientist guaranteed time (GT) program a deep homogeneous \xmm\ survey of 
\m33\ with a sensitivity of $10^{35}$\,erg~s$^{-1}$ in the 0.5--10 keV
band, a factor of ten deeper than previous surveys. The survey consisted of a raster of 15 
pointings of about 10 ks each. The directions were selected in a way that each
area within the \m33\ optical D$_{25}$ extent was covered by the \xmm\ EPIC 
detectors at least three times with the medium filter. The survey was complemented by an 
additional AO2 program, initiated
to fill observation gaps due to high detector background, or, because the 
thick filter was used during the GT observations.
First results of this survey were presented in \citet{2003AN....324...85P}.

Two \m33\ sources are known for their outstanding X-ray properties 
\citep{1989ApJ...336..140P}.
The brightest source \citep[X--8 in the nomenclature of][ luminosity of  
$10^{39 - 40}$\,erg~s$^{-1}$]{1981ApJ...246L..61L} 
is the most luminous X-ray source in the Local Group
of galaxies and coincides with the optical center of \m33. Its time variability  
\citep{1997ApJ...490L..47D}, its point-like nature
seen by ROSAT HRI \citep{2000immm.proc..149P} and \chandra\  
\citep{2002MNRAS.336..901D}
and its X-ray spectrum best described by an absorbed power law plus disk blackbody model 
\cite[e.g. ][]{2001ncxa.conf..300E,2003ApJ...583..758L,fosch2003},
point towards a black hole
XRB. A possible periodicity of 106 days was not
confirmed in later observations \citep[see ][]{2001A&A...368..420P}. The second source (\x7)
is an eclipsing high mass XRB  (HMXB) with a binary period of 3.45 d and
possible 0.31 s pulsations, discovered in ROSAT PSPC and HRI observations 
\citep{1993ApJ...418L..67S,1997ApJ...490L..47D,
1999MNRAS.302..731D,1997AJ....113..618L}. 
\citet{px7} analyzed \xmm\ and \chandra\  data and determined 
improved binary parameters, 
however, they could not confirm the proposed pulsations. In a special 
analysis of DIRECT data, they identified an O7I star of 18.89 mag in V, as the optical 
counterpart, which shows the ellipsoidal heating light curve of a HMXB with the X--7
binary ephemeris. X--7 was the most distant eclipsing XRB until the
detection of another such source in \xmm\ and \chandra\ observations of the 
starburst galaxy \n253, which is located at more than three times  the distance of \m33\ 
\citep{2003A&A...402..457P}.

Here we present merged medium and thin filter images for the three EPIC instruments, in five energy
bands, using only times of low background from our \xmm\ survey of \m33, as well
as a source catalogue and source population study based on these images. 
An analysis of the individual pointings with less stringent background rejection
and using all filters is in progress, and will result in a source catalogue 
with improved positions (taking care of individual pointing offsets), 
that will also provide information on source variability 
and spectral characteristics.

\section{Observations and data analysis}
\begin{table}
\begin{center}
\caption[]{Count rate to energy conversion factor for different spectral models for EPIC instruments in energy band 1 to 5 (B1--B5).
           Models are a power law with photon index of 1.7 (PL), a thin thermal model with temperature of 
	   1 keV (TT), and a black body with a temperature of 30 eV (BB).
	   All models assume galactic foreground absorption of 6\hcm{20} \citep{1992ApJS...79...77S}.}
\begin{tabular}{lrrrrrr}
\hline\noalign{\smallskip}
\hline\noalign{\smallskip}
\multicolumn{1}{l}{Detector} & \multicolumn{1}{c}{Spec} &\multicolumn{1}{c}{B1} &
\multicolumn{1}{c}{B2} & \multicolumn{1}{c}{B3} & 
\multicolumn{1}{c}{B4} & \multicolumn{1}{c}{B5}  \\ 
\noalign{\smallskip}
& & \multicolumn{5}{c}{($10^{-12}$ erg cm$^{-2}$ ct$^{-1}$)} \\
\noalign{\smallskip}\hline\noalign{\smallskip}
EPIC PN & PL & 1.03 & 1.10 & 1.71 & 4.9 & 21. \\
        & TT & 1.03 & 1.14 & 1.55 & 4.0 &  \\
        & BB & 1.09 & 0.97 & 1.30 &     &      \\
EPIC MOS & PL & 6.2 & 4.9 & 4.8 & 12.9 & 66. \\
        & TT & 6.0 & 4.5 & 4.6 & 11.0 &  \\
        & BB & 10.1 & 4.8 & 4.2 &    &      \\
\noalign{\smallskip}
\hline
\noalign{\smallskip}
\end{tabular}
\label{ecf}
\end{center}
\end{table}

\begin{figure}
    \caption[]{\xmm\ EPIC ``PN equivalent" 
    exposure map combining all low background observation
    times of the fields of the \m33 raster. The outer contour indicates 0 ks, 
    gray-scale steps every 10 ks. The maximum PN equivalent exposure (see text) 
    is 89 ks in the center
    area of \m33. The optical extent of \m33 is marked by the white 
    inclination-corrected D$_{25}$ 
    ellipse. The center of the fields of our raster (see
    Table~\ref{observations}) are indicated and numbered.
    \label{expo}}
\end{figure}

Table~\ref{observations} summarizes the \xmm\ \citep{2001A&A...365L...1J} 
EPIC \citep{2001A&A...365L..18S,2001A&A...365L..27T} observations. For each
observation we give
the field number of our \m33\ raster (col. 1), the observation
identification (2), date (3), pointing direction (4,5), as well as instrument
mode, filter and exposure time after screening for high background for EPIC PN
(6,7), MOS1 (8,9), and MOS2 (10,11). Observation 0102640801 was affected by a
satellite ``slew failure" and only the EPIC MOS instruments got 
some exposure time.

In the \xmm\ observations the EPIC PN and MOS instruments 
were mostly operated in the 
full frame mode resulting in  a time
resolution of 73.4 ms and 2.6 s, respectively. Only for four early observations
the PN detector was operated in the extended full
frame mode (time resolution 200 ms). During observation 0102640101 the MOS
detectors were operated in the small window mode (0.3~s time resolution for the
inner CCDs to avoid pile-up for the bright source X$-$8).
The medium filter was in front of the EPIC cameras in all but the first two 
observations which were performed with the thick filter (0102640401, 0102641201) 
and observation 0102640301 for which the MOS2 detector was operated with the thin 
filter. For creating the merged images and for source detection, we only used medium
and thin filter observations.

The data analysis was performed using tools in the \xmm\ Science Analysis System
(SAS) v5.4.1 and some later
versions from the development area as specially mentioned, 
EXSAS/MIDAS 1.2/1.4, and 
FTOOLS v5.2 software packages, the imaging application DS9 v2.3b1 and the spectral 
analysis software XSPEC v11.2.   

For MOS, we used SAS task {\tt emchain} with task {\tt emevent} v7.10 that removes flickering pixels. 
We carefully screened the event files for bad CCD pixels, remaining after the
standard processing. 
To create a homogeneous combined image with similar background level for all
fields we had to carefully screen the data for high background times. 
In most observations the background light curves at all energies follow
the high energy (7-15 keV) background light curves provided by the SAS 
tasks {\tt epchain} and {\tt emchain} and allowed us to efficiently screen the data. 
In some observations, however, the low energy background (below 1 keV) was
higher during high energy selected good time intervals (GTIs) and had to be 
used for additional screening.
GTIs were determined from the higher signal-to-noise PN light
curves and also used for the MOS cameras. Outside the PN time coverage, 
GTIs were determined from the combined MOS light curves. 
The corresponding low background times for the individual observations are
listed for the PN and MOS cameras in Table~\ref{observations}. 
The combined ``PN equivalent" vignetted EPIC exposure (1-2 keV) is
significantly above  20~ks within most of the optical extent
of \m33 (indicated by the inclination-corrected $D_{25}$ ellipse 
in Fig.~\ref{expo}). Only in the NW and
S the exposure does not reach this limit. For the ``PN equivalent" EPIC exposure 
map we added MOS1 and MOS2 exposure maps (scaled by a factor of 0.36 as expected
for a 5 keV thermal bremsstrahlung spectrum with an absorption of 6\hcm{20}, as
expected for XRBs in \m33) to the PN exposure.
   
For the analysis we used five energy bands: (0.2--0.5) keV, (0.5--1.0) keV,
(1.0--2.0) keV, (2.0--4.5) keV, and (4.5--12) keV as band 1 to 5. We
intentionally split the 0.5--2.0 keV band used in the 1XMM XMM-Newton Serendipitous Source 
Catalogue\footnote{Prepared by the XMM-Newton Survey Science Centre Consortium 
({\tt http://xmmssc-www.star.le.ac.uk/})} to get, on average, a more homogeneous
distribution of the source counts to the energy bands, which leads to a better
spread of the hardness ratio values and allows effective source classification
(see Sect. 6).

For PN we
selected only ``singles" (PATTERN=0) in band 1, for the other bands ``singles and
doubles" (PATTERN$\le$4). For MOS we used ``singles" to ``quadruples" 
(PATTERN$\le$12). To avoid background variability over the PN images we
omitted the energy range (7.2--9.2) keV in band 5 where strong fluorescence lines
cause higher background in the outer detector area \citep{2004SPIE.5165..112F}.
To convert source count rates in the individual bands to fluxes we calculated 
count rate to energy conversion factors (ECF) for PN and MOS medium filter 
observations using W3PIMMS and assuming the 
same spectrum as for the first \xmm\ source catalogue, i.e. a 
power law spectrum with photon index 1.7 absorbed by the galactic foreground 
column of 6\hcm{20} (Table~\ref{ecf}). The assumed spectrum represents hard 
sources like XRBs or AGN. We also give ECF values for a typical SNR   
(absorbed 1 keV thin thermal) and SSS (absorbed 30 eV black body) in \m33.
Table~\ref{ecf} demonstrates that for EPIC PN, ECF values only vary by about 20\% 
for the different spectra in bands 1 to 4. The same is true for EPIC MOS, with 
the exception of the 
black body ECF in the 0.2--0.5 keV band where the MOS sensitivity is much lower.
For both PN and MOS, the sensitivity is strongly reduced at energies above 4.5 keV.
On the other hand, background count rates are similar in the hard band. Therefore,
only bright hard sources are detected in this band. 

To classify the source spectra we computed four hardness ratios from the source
count rates. These hardness ratios are defined as 
HRi = (B$_{i+1}$ - B$_{i}$) \/(B$_{i+1}$ + B$_{i}$) , for i = 1 to 4, where
B$_{i}$ denotes count rates in band i, as defined above. 
In the standard source detection products, 
hardness ratios and count rates in individual energy bands are not combined for
all instruments. To improve the statistics, we have combined the count rates (and fluxes) 
from all three EPIC instruments and have then derived hardness ratios for them. 
We are aware that these 
products have some jitter in their meaning, as the relative integration times for 
individual sources in the EPIC instruments differ slightly. Due to the better 
signal in the combined products, they are still very valuable for source 
classification. For investigations needing accurate calibration, values derived
for each instrument separately should be used.

As discussed above,  band 5 suffers from the combined effect of lower sensitivity 
in  both PN and MOS, and a large contamination from the hard background, 
reducing the significance of any detection due to the much higher noise.  We 
therefore give total count rates and errors derived from all instruments in the 
``XID" band that comprises bands 1--4, and derive  fluxes in the corresponding 
0.2-4.5 keV range.  A larger band would artificially increase the overall noise 
of all but extremely hard sources.  For very soft sources, or for sources  
detected only in one of the instruments ($e.g.$ soft sources in PN), even with 
this choice the quoted errors are overestimated, however the number of sources 
of this nature is small.

For PN, MOS1 and MOS2 we created images,
background images, exposure maps (without and with vignetting correction) and
masked them for the acceptable detector area in each of the five energy bands.
For PN, the background maps contain the contribution from the ``out of time (OOT)"
events (parameter {\tt withootset=true} in task {\tt esplinemap}). The images
of the individual observations were then merged for source detection into 
images covering the full \m33\ field. The maximum
vignetted exposure was 44.3 ks, 58.6 ks and 58.7 ks in the 1--2 keV band 
for the PN, MOS1 and MOS2 cameras, respectively, and was nearly the same for
bands 1 to 4. In the hardest band, the maximum exposure is smaller by 30\% due to the larger 
vignetting effects. 
To allow us an easy merging of images, we calculated for the events of all 
observations projected sky coordinates (X,Y) with respect to the center reference position 
RA=01$^{\rm h}$33$^{\rm m}$50\fs4, DEC=$+$30\degr39\arcmin36\arcsec (J2000). 
At the borders, due to the different field of view of the EPIC instruments, 
the merged images for PN, MOS1 and MOS2 do not fully overlap. Nevertheless, 
we used the full area for source detection.

\begin{table*}
\scriptsize
\caption[]{X-ray source catalogue of the \xmm\ EPIC \m33\ raster observation.}
\begin{tabular}{rrrrrrrrrrrr}
\hline\noalign{\smallskip}
\hline\noalign{\smallskip}
\multicolumn{1}{c}{Src} & 
\multicolumn{1}{c}{RA(J2000)} &
\multicolumn{1}{c}{DEC(J2000)} &
\multicolumn{1}{c}{r$_{1\sigma}$} &
\multicolumn{1}{c}{ML$_{\rm exi}$} & 
\multicolumn{1}{c}{Count rate$^*$} &
\multicolumn{1}{c}{Flux$^*$}  &
\multicolumn{1}{c}{HR1} &
\multicolumn{1}{c}{HR2} &
\multicolumn{1}{c}{HR3} &
\multicolumn{1}{c}{HR4} &
\multicolumn{1}{c}{Val} \\ 
\multicolumn{1}{c}{} & 
\multicolumn{1}{c}{(h mm ss.ss)} &
\multicolumn{1}{c}{(+dd mm ss.s)} &
\multicolumn{1}{c}{(\arcsec)} &
\multicolumn{1}{c}{} & 
\multicolumn{1}{c}{(ct s$^{-1}$)} &
\multicolumn{1}{c}{(erg cm$^{-2}$ s$^{-1}$)} &
\multicolumn{1}{c}{} \\ 
\multicolumn{1}{c}{(1)} & 
\multicolumn{1}{c}{(2)} &
\multicolumn{1}{c}{(3)} &
\multicolumn{1}{c}{(4)} &
\multicolumn{1}{c}{(5)} & 
\multicolumn{1}{c}{(6)} &
\multicolumn{1}{c}{(7)} &
\multicolumn{1}{c}{(8)} &
\multicolumn{1}{c}{(9)} &
\multicolumn{1}{c}{(10)} &
\multicolumn{1}{c}{(11)} &
\multicolumn{1}{c}{(12)} \\ 
\noalign{\smallskip}\hline\noalign{\smallskip}
  1 & 1 31 42.93 &  +30 23 29.5 &  1.78 &  3.41e+01 & 1.18e-02$\pm$2.2e-03 & 1.55e-14$\pm$4.0e-15& 0.44$\pm$0.21 & -0.39$\pm$0.21 & -0.84$\pm$0.34&	 *$\pm$   * &TFF \\
  2 & 1 31 46.89 &  +30 22 01.3 &  1.03 &  1.43e+02 & 3.86e-02$\pm$4.1e-03 & 4.17e-14$\pm$5.7e-15& 0.44$\pm$0.11 & -0.17$\pm$0.12 & -0.75$\pm$0.14&	 *$\pm$   * &TTF \\
  3 & 1 31 54.90 &  +30 29 54.5 &  1.86 &  2.48e+01 & 7.15e-03$\pm$1.6e-03 & 6.33e-15$\pm$1.7e-15& 0.37$\pm$0.21 & -0.64$\pm$0.26 &	*$\pm$   *&	 *$\pm$   * &TTT \\
  4 & 1 31 55.12 &  +30 28 56.9 &  1.62 &  1.49e+01 & 6.37e-03$\pm$1.7e-03 & 7.72e-15$\pm$2.7e-15&    *$\pm$   * &  0.31$\pm$0.31 & -0.11$\pm$0.29&   0.20$\pm$0.35 &TTT \\
  5 & 1 31 59.18 &  +30 35 06.9 &  1.60 &  1.58e+01 & 5.00e-03$\pm$1.0e-03 & 6.96e-15$\pm$1.7e-15& 0.38$\pm$0.27 & -0.37$\pm$0.27 &  0.37$\pm$0.28&  -1.00$\pm$0.90 &TTT \\
...\\
408 & 1 36 01.07 &  +30 51 28.0 &  1.10 &  1.10e+02 & 2.52e-02$\pm$2.7e-03 & 2.72e-14$\pm$4.1e-15& 0.82$\pm$0.13 &  0.06$\pm$0.11 & -0.32$\pm$0.14&  -0.20$\pm$0.31 &TTT \\
\noalign{\smallskip}
\hline
\noalign{\smallskip}
\end{tabular}
\begin{tabular}{rrrrrrrrll}
\hline\noalign{\smallskip}
\hline\noalign{\smallskip}
\multicolumn{1}{c}{XID} & 
\multicolumn{1}{c}{USNO-B1} &
\multicolumn{1}{c}{Mul} &
\multicolumn{1}{c}{$\Delta$} &
\multicolumn{1}{c}{B2} & 
\multicolumn{1}{c}{R2} &
\multicolumn{1}{c}{I2}  &
\multicolumn{1}{c}{log(${\rm f}_{\rm x} \over {\rm f}_{\rm opt}$)} &
\multicolumn{1}{c}{Class} &
\multicolumn{1}{c}{Remarks}\\ 
\multicolumn{1}{c}{} & 
\multicolumn{1}{c}{name} &
\multicolumn{1}{c}{} &
\multicolumn{1}{c}{(\arcsec)} &
\multicolumn{1}{c}{(mag)} & 
\multicolumn{1}{c}{(mag)} &
\multicolumn{1}{c}{(mag)} \\ 
\multicolumn{1}{c}{(13)} & 
\multicolumn{1}{c}{(14)} &
\multicolumn{1}{c}{(15)} &
\multicolumn{1}{c}{(16)} &
\multicolumn{1}{c}{(17)} & 
\multicolumn{1}{c}{(18)} &
\multicolumn{1}{c}{(19)} &
\multicolumn{1}{c}{(20)} &
\multicolumn{1}{c}{(21)} &
\multicolumn{1}{c}{(22)} \\ 
\noalign{\smallskip}\hline\noalign{\smallskip}
      		      & 1203-0020312 & 1& 3.1& 15.8& 13.5& 11.3&  -2.6&  fg Star    &W UMa Var(SIM)     \\  
LCB2(Var(flare)),HP8  & 2293-01403-1 & 1& 2.7& 11.1& 10.8& 10.6&  -3.6&  fg Star    &beta Lyrae EB(SIM) \\
		      & 2293-01131-1 & 1& 2.7& 12.4& 11.3& 10.9&  -4.1&  $<{\rm fg Star}>$  & $<{\rm G8}>$	 \\	  
		      &		& *&   *&    *&    *&	 *&	*& $<{\rm hard}>$ &		   \\	  
		      &		& *&   *&    *&    *&	 *&	*&	       &		   \\	  
...\\
         	      & 	     & *&   *&    *&	*&    *&     *& $<{\rm hard}>$ &	\\
\noalign{\smallskip}
\hline
\noalign{\smallskip}
\end{tabular}

\label{master}
Notes:\\
$^*$: in XID band (0.2--4.5 keV)       \\
References:\\
in XID: X-n: \citet{1981ApJ...246L..61L}, \citet{1988ApJ...329.1037T}, SBn: \citet{1995ApJ...441..568S}, 
LCBn: \citet{1996ApJ...466..750L}, DCL97: \citet{1997ApJ...490L..47D}, DCL99: \citet{1999MNRAS.302..731D}, 
HPn: Haberl \& Pietsch (2001)\\
in remarks: SIM: SIMBAD database, NED: NASA Extragalactic Database, NVSS: \citet{1998AJ....115.1693C},
CS82: \citet{1982ApJ...253L..13C},
[MC83]25: \citet{1983ApJ...273..576M}, IFM*: \citet{1993ApJS...89...85I}, GKLn: \citet{1998ApJS..117...89G}, 
GDKn: \citet{1999ApJS..120..247G}, IV74: \citet{1974A&A....32..363I}, VBW78: \citet{1978A&A....62...51V}, 
M98: \citet{1998ApJ...501..153M}, MCM2001: \citet{2001A&A...367..498M}, MBH96: \citet{1996ApJ...469..629M}, 
PMM2004: \citet{px7}, GM94: \citet{1994PASP..106..376G}, FSZ: \citet{1997BSAO...43..133F}, 
D33: \citet{2001AJ....121..870M},\citet{2001AJ....121.2032M}, RW93: \citet{1993AJ....105..499R}, 
VHK: \citet{1926ApJ....63..236H}, CGC97: \citet{1997AJ....114.2353C}, BCLMP: \citet{1974A&A....37...33B}
\normalsize
\end{table*}

\begin{table*}
\scriptsize
\caption[]{X-ray source catalogue of the \xmm\ EPIC \m33\ raster observation.}
\begin{tabular}{rrrrrrrrrr}
\hline\noalign{\smallskip}
\hline\noalign{\smallskip}
\multicolumn{1}{c}{Src} & 
\multicolumn{1}{c}{Val} &
\multicolumn{8}{l}{EPIC PN parameters} \\
\multicolumn{1}{c}{} &
\multicolumn{1}{c}{} &
\multicolumn{1}{c}{Exposure} &
\multicolumn{1}{c}{ML$_{\rm exi}$} & 
\multicolumn{1}{c}{Count rate$^*$} &
\multicolumn{1}{c}{Flux$^*$}  &
\multicolumn{1}{c}{HR1} &
\multicolumn{1}{c}{HR2} &
\multicolumn{1}{c}{HR3} &
\multicolumn{1}{c}{HR4} \\ 
\multicolumn{1}{c}{} & 
\multicolumn{1}{c}{} &
\multicolumn{1}{c}{(ks)} &
\multicolumn{1}{c}{} & 
\multicolumn{1}{c}{(ct s$^{-1}$)} &
\multicolumn{1}{c}{(erg cm$^{-2}$ s$^{-1}$)} \\ 
\multicolumn{1}{c}{(1)} & 
\multicolumn{1}{c}{(2)} &
\multicolumn{1}{c}{(3)} &
\multicolumn{1}{c}{(4)} &
\multicolumn{1}{c}{(5)} & 
\multicolumn{1}{c}{(6)} &
\multicolumn{1}{c}{(7)} &
\multicolumn{1}{c}{(8)} &
\multicolumn{1}{c}{(9)} &
\multicolumn{1}{c}{(10)} \\ 
\noalign{\smallskip}\hline\noalign{\smallskip}
  1 & TFF &  2.7 & 3.41e+01 & 1.18e-02$\pm$2.2e-03 & 1.55e-14$\pm$4.0e-15 &  0.44$\pm$0.21 & -0.39$\pm$0.21 & -0.84$\pm$0.34 &     *$\pm$  *  \\
  2 & TTF &  2.7 & 1.32e+02 & 3.30e-02$\pm$3.7e-03 & 4.94e-14$\pm$7.2e-15 &  0.38$\pm$0.12 & -0.14$\pm$0.13 & -0.71$\pm$0.16 &     *$\pm$  *  \\
  3 & TTT &  3.8 & 1.04e+01 & 4.40e-03$\pm$1.4e-03 & 5.91e-15$\pm$2.4e-15 &  0.65$\pm$0.33 & -0.59$\pm$0.32 &	  *$\pm$   * &     *$\pm$  *  \\
  4 & TTT &  3.8 & 1.46e+01 & 4.42e-03$\pm$1.4e-03 & 1.05e-14$\pm$4.1e-15 &	*$\pm$   * &  0.27$\pm$0.34 & -0.32$\pm$0.36 &     *$\pm$  *  \\
  5 & TTT &  8.4 & 1.57e+01 & 3.50e-03$\pm$8.5e-04 & 9.23e-15$\pm$2.8e-15 &  0.43$\pm$0.32 & -0.45$\pm$0.31 &  0.48$\pm$0.31 & -1.00$\pm$1.02 \\
... \\
408 & TTT &  4.1 & 5.06e+01 & 1.17e-02$\pm$1.7e-03 & 2.14e-14$\pm$4.7e-15 &  1.00$\pm$0.08 & -0.07$\pm$0.14 & -0.54$\pm$0.22 &     *$\pm$  *  \\
\noalign{\smallskip}
\hline
\noalign{\smallskip}
&&\multicolumn{8}{l}{EPIC MOS1 parameters} \\
&&\multicolumn{1}{c}{Exposure} &
\multicolumn{1}{c}{ML$_{\rm exi}$} & 
\multicolumn{1}{c}{Count rate$^*$} &
\multicolumn{1}{c}{Flux$^*$}  &
\multicolumn{1}{c}{HR1} &
\multicolumn{1}{c}{HR2} &
\multicolumn{1}{c}{HR3} &
\multicolumn{1}{c}{HR4} \\ 
&&\multicolumn{1}{c}{(ks)} &
\multicolumn{1}{c}{} & 
\multicolumn{1}{c}{(ct s$^{-1}$)} &
\multicolumn{1}{c}{(erg cm$^{-2}$ s$^{-1}$)} \\ 
&&\multicolumn{1}{c}{(11)} & 
\multicolumn{1}{c}{(12)} &
\multicolumn{1}{c}{(13)} &
\multicolumn{1}{c}{(14)} &
\multicolumn{1}{c}{(15)} & 
\multicolumn{1}{c}{(16)} &
\multicolumn{1}{c}{(17)} &
\multicolumn{1}{c}{(18)} \\ 
\noalign{\smallskip}\hline\noalign{\smallskip}
&&   *&	      *&	 *$\pm$      *& 	*$\pm$      *&      *$\pm$   *&      *$\pm$   * &     *$\pm$   *&      *$\pm$   *  \\
&& 2.8&  1.42e+01&  5.61e-03$\pm$1.7e-03&  2.78e-14$\pm$9.6e-15&   0.76$\pm$0.30&  -0.28$\pm$0.30 &     *$\pm$   *&      *$\pm$   *  \\
&& 9.5&  2.64e-01&  4.38e-04$\pm$4.5e-04&  2.60e-15$\pm$3.2e-15&      *$\pm$   *&      *$\pm$   * &     *$\pm$   *&      *$\pm$   *  \\
&& 4.9&  1.38e-01&  3.15e-04$\pm$4.2e-04&  2.57e-15$\pm$4.1e-15&      *$\pm$   *&      *$\pm$   * &     *$\pm$   *&      *$\pm$   *  \\
&&10.7&  1.65e+00&  5.80e-04$\pm$3.2e-04&  5.14e-15$\pm$3.2e-15&      *$\pm$   *&      *$\pm$   * &     *$\pm$   *&      *$\pm$   *  \\
&&... \\
&& 4.1&  3.84e+01&  7.34e-03$\pm$1.5e-03&  5.23e-14$\pm$1.3e-14&   1.00$\pm$0.34&   0.23$\pm$0.22 & -0.23$\pm$0.23&  -0.02$\pm$0.39  \\
\noalign{\smallskip}
\hline
\noalign{\smallskip}
&&\multicolumn{8}{l}{EPIC MOS2 parameters} \\
&&\multicolumn{1}{c}{Exposure} &
\multicolumn{1}{c}{ML$_{\rm exi}$} & 
\multicolumn{1}{c}{Count rate$^*$} &
\multicolumn{1}{c}{Flux$^*$}  &
\multicolumn{1}{c}{HR1} &
\multicolumn{1}{c}{HR2} &
\multicolumn{1}{c}{HR3} &
\multicolumn{1}{c}{HR4} \\ 
&&\multicolumn{1}{c}{(ks)} &
\multicolumn{1}{c}{} & 
\multicolumn{1}{c}{(ct s$^{-1}$)} &
\multicolumn{1}{c}{(erg cm$^{-2}$ s$^{-1}$)} \\ 
&&\multicolumn{1}{c}{(19)} & 
\multicolumn{1}{c}{(20)} &
\multicolumn{1}{c}{(21)} &
\multicolumn{1}{c}{(22)} &
\multicolumn{1}{c}{(23)} & 
\multicolumn{1}{c}{(24)} &
\multicolumn{1}{c}{(25)} &
\multicolumn{1}{c}{(26)} \\ 
\noalign{\smallskip}\hline\noalign{\smallskip}
&&   *&	      *&	 *$\pm$      *& 	*$\pm$      *&      *$\pm$   *&      *$\pm$   * &     *$\pm$   *&      *$\pm$   *  \\
&&   *&	      *&	 *$\pm$      *& 	*$\pm$      *&      *$\pm$   *&      *$\pm$   * &     *$\pm$   *&      *$\pm$   *  \\
&& 9.4&  1.84e+01&  2.32e-03$\pm$6.1e-04&  1.26e-14$\pm$3.8e-15&   0.01$\pm$0.26&  -0.72$\pm$0.31 &     *$\pm$   *&      *$\pm$   *  \\
&& 4.7&  4.22e+00&  1.64e-03$\pm$7.6e-04&  1.54e-14$\pm$7.4e-15&      *$\pm$   *&      *$\pm$   * &     *$\pm$   *&      *$\pm$   *  \\
&&10.4&  1.83e+00&  9.21e-04$\pm$4.5e-04&  5.84e-15$\pm$3.1e-15&      *$\pm$   *&      *$\pm$   * &     *$\pm$   *&      *$\pm$   *  \\
&&... \\
&& 4.0&  2.35e+01&  6.09e-03$\pm$1.5e-03&  4.43e-14$\pm$1.3e-14&      *$\pm$   *&   0.17$\pm$0.32 & -0.10$\pm$0.33&  -0.38$\pm$0.58  \\
\noalign{\smallskip}
\hline
\noalign{\smallskip}
\end{tabular}

\label{master_pn_mos}
Notes:\\
$^*$: in XID band (0.2--4.5 keV)       \\
\normalsize
\end{table*}

\section{Source catalogue}
We searched for sources in these merged images using
simultaneously $5\times 3$ images (5 energy bands and PN, MOS1 and MOS2 camera). 
A preliminary
source list created with the task {\tt boxdetect} with a low likelihood threshold 
was used as starting point for the task {\tt emldetect v4.31}. As photons
collected with different off-axis angles contribute to the source counts in the
merged image, we prepared a specific point spread function (PSF) calibration 
file. It assumed a constant PSF over the total field of view that was 
degraded as if 6\arcmin\ off-axis. 
This is the most degraded but still ``circular" PSF in 
the EPIC calibration files and a first approximation for the PSF when source counts 
are collected at many different off-axis angles as in the \m33\ raster observation. 
To resolve
sources that overlap due to the PSF, we used parameter {\tt multisourcefit=4}.
 
A detection probability above 4$\sigma$ is
equivalent to a single band limiting detection likelihood of 10. This threshold
has been  successfully used by HP01 for the ROSAT PSPC detections (see Sect. 5). 
In the case of simultaneous
detection with a large number of images this equivalence is no longer valid due
to the small number of counts in the individual images contributing to a source
at the detection threshold. We have established for the task {\tt emldetect} 
that for the case of simultaneously using 15 images, a detection 
probability above 4$\sigma$ corresponds to a combined 
likelihood of 7. This threshold was used to select sources for the catalogue.
The detection procedure
produced acceptable results as can be checked from the overlay of the
detected sources onto the smoothed X-ray 
images (see Sect.~4).

Two emission peaks were resolved into several sources with no sign of 
multiplicity from the
smoothed images and are better described as extended. In a detection run including extent determination 
 we found that these two sources 
(\#125 and \#299) can be characterized by a Gaussian with FWHM of
18\farcs0 and 16\farcs2, respectively, and the extent likelihood is high. In total we detected 408 sources in the
field.

The source parameters are summarized in Tables~\ref{master} (EPIC combined products) and
\ref{master_pn_mos} (products for EPIC PN, MOS1 and MOS2, separately). 
For both tables we only show the first five and the last row in the
paper. The full tables are available in electronic form at the CDS
{\bf (Editors should add URL address here...)}. 

Table~\ref{master} 
gives the source number (col. 1), position (2 and 3) with $1\sigma$
uncertainty radius (4), likelihood of existence (5), integrated PN,
MOS1 and MOS2 count rate (6) and flux (7) in the 0.2--4.5 keV band, 
and hardness ratios (8--11). 
Hardness ratios are calculated only for sources for which at 
least one of the two band count rates has a significance greater than $2\sigma$.  
Errors are the properly combined statistical errors in each band and can
extend beyond the range of allowed values of hardness ratios as defined previously 
(-1.0 to 1.0). 
The EPIC instruments contributing to the source 
detection, are indicated in the ``Val" parameter (col. 12, first character for 
PN, second MOS1, third MOS2) as ``T", if inside the field of view (FOV), or ``F", 
if outside of FOV. 
There are only nine sources at the periphery of the FOV where only part of the EPIC 
instruments contribute. The positional error does not include intrinsic systematic errors 
which, according to the Users Guide of the 1XMM XMM-Newton Serendipitous Source 
Catalogue, amount to 1\farcs5 and should be 
quadratically added to the statistical errors in col. 4. 
In the remaining columns of Table~\ref{master}, we give cross correlation information,
which is further described in Sect.~5. We only want to mention here, that we used the 
foreground stars and candidates, to verify the assumed  source position errors. All 35
foreground stars and candidates are located within the $3\sigma$ statistical plus systematic 
positional error given above.

The faintest sources detected have a flux of 1.0\ergcm{-15}, the brightest source (X-8)
1.1\ergcm{-11} in the 0.2--4.5 keV band. This corresponds to an absorbed luminosity range 
in \m33\ of 7.6\ergs{34} to 8.4\ergs{38}.

Table~\ref{master_pn_mos} repeats source number and Val parameter of Table~\ref{master}
(cols. 1 and 2). It then gives for EPIC PN exposure (3), source existence 
likelihood (4),  count rate (5) and flux (6) in the 0.2--4.5 keV band,
and hardness ratios (7--10). Columns 11 to 18 and 19 to 26 give the same information 
corresponding to cols. 3 to 10, but now for the EPIC MOS1 and MOS2 instruments. Hardness ratios 
were again screened as in  Table~\ref{master}. From the comparison of the hardness ratios
in  Table~\ref{master} and \ref{master_pn_mos} it is clear that combining the instrument 
count rate information, yielded significantly more hardness ratios above the chosen 
significance threshold.
 
\section{Images}
\begin{figure*}
    \caption[]{
Logarithmically-scaled \xmm\ EPIC low background image of the \m33\ 
medium and thin filter observations combining PN and MOS1 and MOS2 
cameras in the (0.2--4.5)~keV band. 
The data are smoothed with a Gaussian
of FWHM 20\arcsec\ which corresponds to the average point spread function of 
the combined observations with different off-axis angles. The image is 
corrected for un-vignetted exposure and masked for exposure above 14 ks. 
Contours are at $(0.8, 1.2, 2.0, 3.6)\times 10^{-5}$ ct s$^{-1}$. 
Sources from the catalogue, RA, DEC (J2000.0) coordinates, and the 
optical D$_{25}$ ellipse of \m33\ are marked.
}
    \label{imax} 
\end{figure*}
\begin{figure*}
     \caption[]{
     \xmm\ EPIC \m33\ images: band 1--4 (upper left),
     band 1 (upper right),
     band 2 (lower left), and 
     band 3 (lower right).
}
    \label{imax_inner} 
\end{figure*}
\begin{figure*}
\addtocounter{figure}{-1}
     \caption[]{(continued)
     \xmm\ EPIC \m33\ images:  
     band 4 (left) and
     band 5 (right). The images are smoothed with a Gaussian
     of FWHM 20\arcsec, corrected for un-vignetted exposure and masked for 
     exposure above 14 ks. Contour levels are 
     $(8, 12, 20, 36)\times 10^{-6}$ ct s$^{-1}$ for the merged band 1--4,
     $(1.5, 2, 3, 5)\times 10^{-6}$ ct s$^{-1}$ for the band 1,
     $(1.7, 2.2, 3.2, 5.2)\times 10^{-6}$ ct s$^{-1}$ for the band 2,
     $(3, 4, 6, 10)\times 10^{-6}$ ct s$^{-1}$ for the band 3,
     $(2.5, 3.5, 5.5, 9.5)\times 10^{-6}$ ct s$^{-1}$ for the band 4, and
     $(5, 6, 8, 12)\times 10^{-6}$ ct s$^{-1}$ for the band 5 images.
     The inclination-corrected optical D$_{25}$ ellipse of \m33\ , the sky orientation, 
     and the sources from the catalogue 
     - including source numbers in the (4.5--12) keV band image - are 
     indicated. 
}
\end{figure*}
\begin{figure*}
  \hfill
  \parbox[b]{55mm}{
    \caption[]{
Logarithmically-scaled, three-colour \xmm\ EPIC low background image of the \m33\ 
medium and thin filter observations combining PN and MOS1 and MOS2 
cameras. Red, green and
blue show respectively the (0.2--1.0)~keV, (1.0-2.0)~keV 
and (2.0--12.0)~keV bands. 
The data in each energy band have been smoothed with a Gaussian
of FWHM 20\arcsec\ which corresponds to the average point spread function of 
the combined observations with different off-axis angles. The image has been 
corrected for un-vignetted exposure and masked for exposure above 14 ks. The 
image scale and the inclination-corrected optical D$_{25}$ ellipse of \m33\ are marked.
}
    \label{rgb}} 
\end{figure*}
For Figs.~\ref{imax},~\ref{imax_inner},~\ref{rgb},
we smoothed the images and un-vignetted exposure maps per band and observation 
(created as described in Sect.~2) with a Gaussian (FWHM of 20\arcsec) and masked 
them for acceptable detector area.
For PN, we subtracted OOT images, that were masked and smoothed in the same way
as the images. We normalized the images with the un-vignetted exposure maps to
avoid an over-correction of the partly un-vignetted detector background.
For the EPIC combined images, we added the images of the individual
cameras scaled according to the background in the individual energy bands. For
the colour image and the broad band image we added images for the individual bands
as needed and masked them to a total un-vignetted exposure of 10 ks. To better
visualize faint structures we add contours. We indicate the detected sources 
by boxes and the optical extent of \m33\ by the inclination-corrected D$_{25}$ ellipse.

The XID band (0.2-4.5 keV) image (Fig.~\ref{imax}) gives an overview on the
sources detected in the \xmm\ analysis. Two of the sources (1 and 7) 
are bright sources detected in areas with low exposure that were masked out for 
the images. In Fig.~\ref{imax_inner} we rotated the images of the individual 
band in a way that the major axis of the D$_{25}$ ellipse is oriented
vertically. We over-plotted the positions of the sources of our catalogue. By
comparing the different energy bands it is clear that many sources only show up
in some of the images, a fact that indicates spectral diversity and is further 
quantified in the different hardness ratios of the sources. This fact can be
visualized even more clearly in the combined EPIC colour image where we coded
the 0.2--1.0 keV band in red, 1.0--2.0 keV band in green and above 2.0 keV in
blue (see Fig.~\ref{rgb}). The image is a demonstration of the colourful X-ray
sky.  SSS, thermal SNRs and foreground stars appear red
or yellow, XRBs, Crab-like SNRs and AGN green to blue. Diffuse emission fills
the inner area of the disk and the southern spiral arm (red, see also 
Fig.~\ref{imax_inner}, band 1--3). A detailed analysis of this emission is 
outside the scope of this paper.

The images reveal that there is additional emission that has not been resolved
into individual sources. This may be due to really diffuse emission or due to sources that did
not reach the detection threshold. Examples are, structures in the area around 
the three bright sources X--9a,b,c (277, 297, 310) in the NE, source X--4 (124) in 
the NW and source X--6 (155) in the SSW. There seems to be a relatively bright
source which was missed by the detection procedure between sources 235 and 258.  
Faint structure appears between sources 55 and 65 and south of source 91. Only
deeper and/or higher resolution observations will allow us to clarify the nature
of this emission.
\section{Cross-correlation with other \m33\ X-ray catalogues}
\begin{table}
\begin{center}
\caption[]{Summary of cross-correlation with the \m33\ ROSAT catalogue of HP01.
ROSAT sources, not detected in our \xmm\ analysis, are arranged according to
their ROSAT detection likelihood LH.}
\begin{tabular}{ll}
\hline\noalign{\smallskip}
\hline\noalign{\smallskip}
\multicolumn{1}{c}{classification} & 
\multicolumn{1}{c}{HP01 source numbers} \\ 
\noalign{\smallskip}\hline\noalign{\smallskip}
outside \xmm\ field & 1,2,3,4,5,6,7,10,11,46,167,171,\\
                    & 176,177,180,181,182,183,184\\
\noalign{\smallskip}
within extended source 125	& 48,50,51\\
\noalign{\smallskip}
resolved by \xmm & 22,29,59,121,172 \\
\noalign{\smallskip}
not detected: LH$<$9 & 14,16,24,35,37,38,65,73,79,85,88,\\
(28 of 34 sources)   & 89,94,105,107,108,116,127,131,\\
		     & 135,138,149,153,155,159,166,175\\
\noalign{\smallskip}
not detected: 9$\le$LH$<$10 & 23,36,43,47,56,72,90,99,100,101,\\
(13 of 19 sources)          & 113,125,142\\
\noalign{\smallskip}
not detected: LH$\ge$10 & 9(LH=29),30(59),41(44),53(13),\\
                       & 68(42),69(11),82(27),83(14),\\
		       & 93(198,near nucleus),97(11),\\
		       & 98(21,near nucleus),119(14,near \\
		       & nucleus),128(18),133(20)\\
\noalign{\smallskip}
\hline
\noalign{\smallskip}
\end{tabular}
\label{rosat}
\end{center}
\end{table}
In this section we discuss the cross-correlation of the \xmm\ 
detected sources with sources
reported in earlier X-ray catalogues. All correlations (together with other
X-ray information like variability, reported in these catalogues, or extent,
detected in this work) 
are indicated in the XID column of Table~\ref{master} (col. 13).

From the 17 \ein\ sources reported in
\citet{1988ApJ...329.1037T} we detected all but the relatively bright  source
X--12 in the northern disk and the less than $3\sigma$ detection X--15 
which was located about 90\arcsec\ to the East of the nucleus. Both of these
sources were also not detected by ROSAT 
\citep[][HP01]{1995ApJ...441..568S,1996ApJ...466..750L}. While one may question
the original existence of X--15, X--12 was detected by \ein\ with high
significance and has to be strongly time variable.

In the correlation with ROSAT detected sources, we concentrated on the catalogue
derived by HP01 from all archival observations of the field, which contains 
184 sources within 50\arcmin\ of the nucleus. It re-detects all 37 sources detected
in the ROSAT PSPC analysis of \citet{1996ApJ...466..750L} and 21 of the 27 
sources from the first ROSAT HRI analysis of the central area of \m33\ by
\citet{1995ApJ...441..568S}. We searched for these low signal-to-noise 
sources (numbers 18, 19, 20, 23,
24, and 27) without success in the more sensitive \xmm\ data. The sources
therefore most likely are not just time variable, but rather false detections.

In Table~\ref{rosat}, we classified sources of HP01 who analyzed all archival 
ROSAT PSPC and HRI observations of the \m33\ field: 19 sources are positioned outside 
the field covered by the \xmm\ observations, three are detected within the
extended emission of \xmm\ source 125, 5 are resolved into multiple components 
in the \xmm\ catalogue. We did not detect 41 out of 51 sources with a ROSAT
detection likelihood below 10.
All these sources originate from the ROSAT 
HRI detections list which, following the experience at the time, was generated
with a likelihood limit of 8 for detection compared to the limit of 10 used for 
PSPC sources. If still at the ROSAT brightness, 
these HRI sources should have been detected by the deeper \xmm\
observations. 
While, in principle, they could all have dimmed by such an amount
that we did not detect them, this seems rather unlikely and probably most of
these sources were spurious detections. The situation is certainly different for
the HP01 sources with likelihood above 10. Many of these sources were detected
with both ROSAT detectors and the non-detection with \xmm\ indicates time variability.
HP01 source 30 was already found to vary by more than a factor of 7.5 during the
ROSAT observations, two other
sources (69 and 82) were classified by HP01 as SSS candidates, which as a
class are known to show strong time variability, and another 
two were classified as 
foreground stars (9 and 133), which might have been flaring during the ROSAT
observations. The HP01 sources close to the nucleus of
M33 (93, 98, and 119) may have been missed in our \xmm\ analysis due to
contamination from the bright nuclear source X--8 or could be spurious
detections due to structures in the ROSAT PSF. One other HP01 source (128) is clearly 
visible in the \xmm\ band 3 and 4 images (see Fig.~\ref{imax_inner}) between \xmm\ 
sources 235 and 258, however it was not picked up by the detection algorithms. 
Besides these special cases, there are 92 clear and four 
questionable (214, 256, 258, 338) \xmm\ and HP01 correlations listed in col. 13 of
Table~\ref{master}. 

\section{Classes of point-like X-ray sources detected in the direction of \m33}
\begin{figure}
   \resizebox{\hsize}{!}{\includegraphics[bb=50 90 355 460,angle=-90,clip]{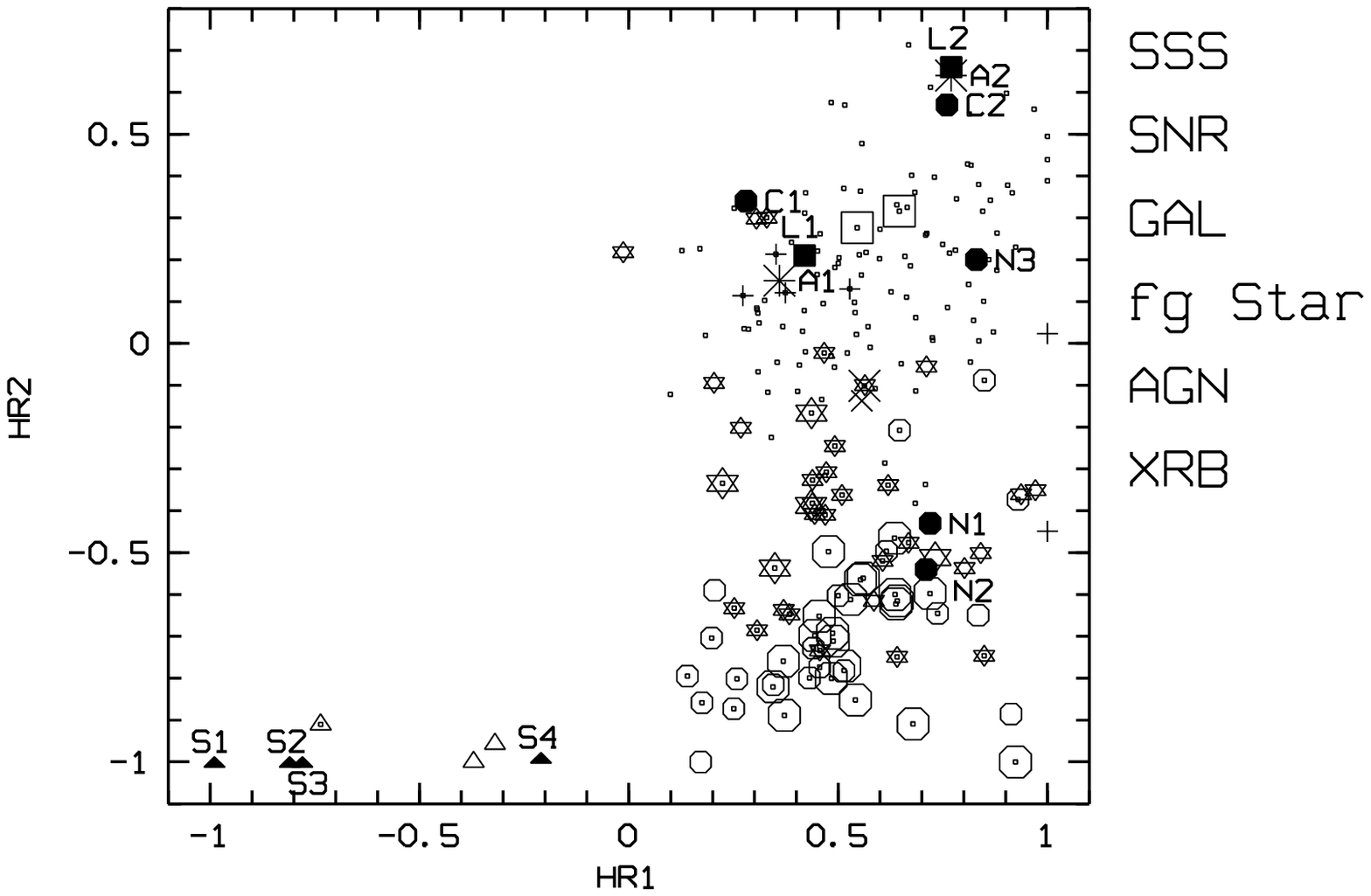},\hskip1.0cm}
   \resizebox{\hsize}{!}{\includegraphics[bb=50 90 340 460,angle=-90,clip]{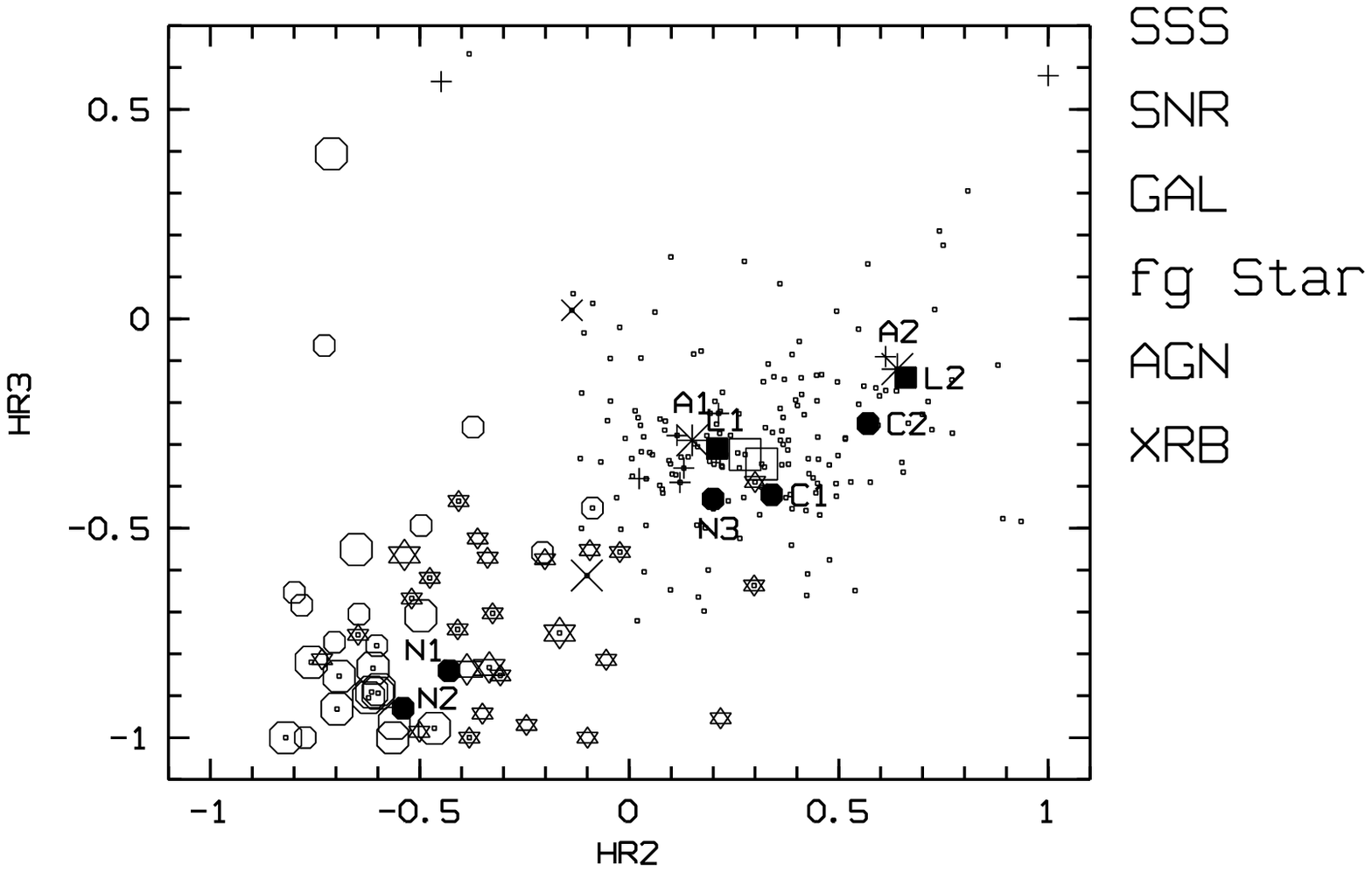},\hskip1.0cm}
   \resizebox{\hsize}{!}{\includegraphics[bb=50 90 355 460,angle=-90,clip]{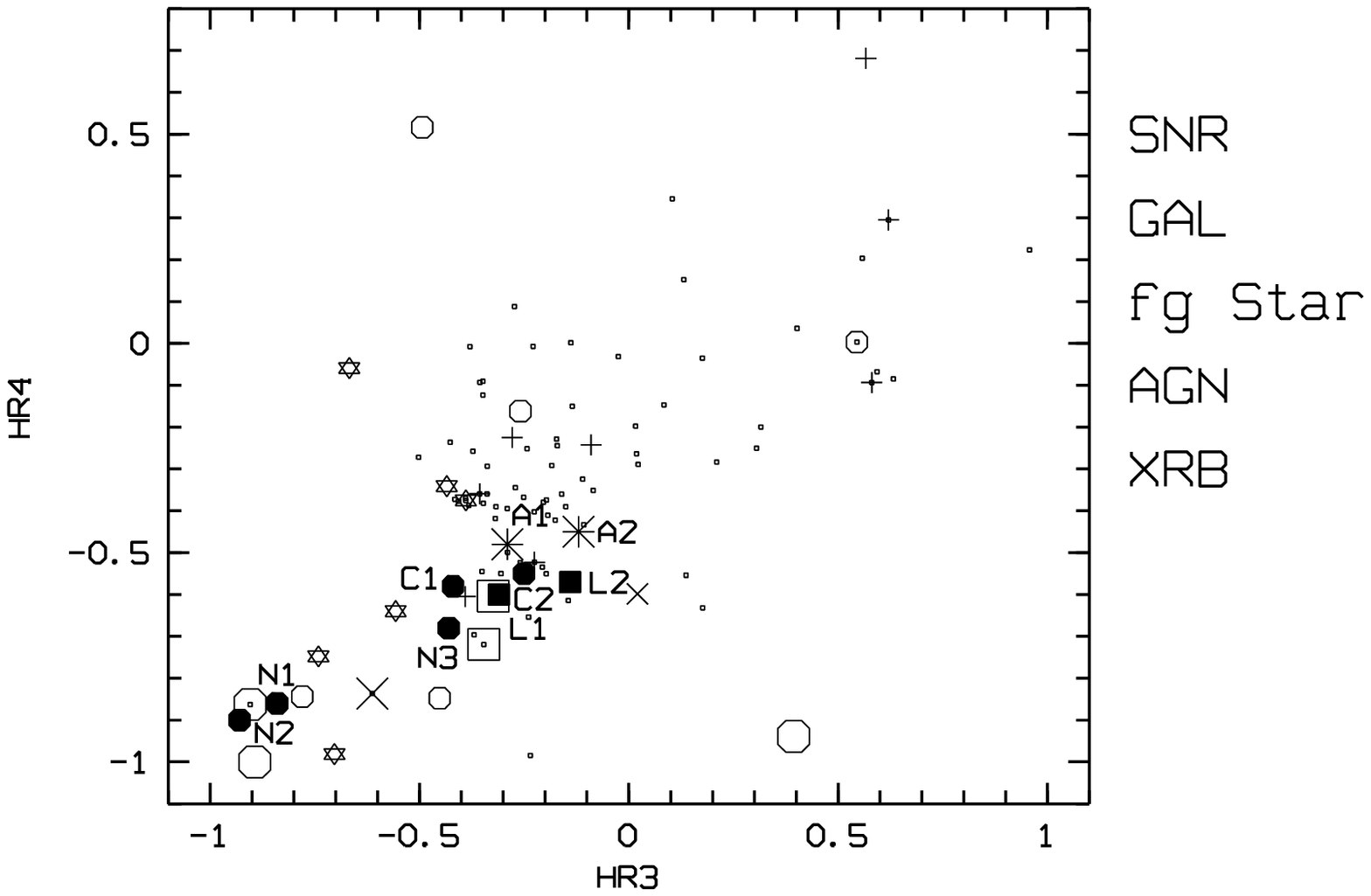},\hskip1.0cm}
     \caption[]{
     Hardness ratios of sources detected by \xmm\ EPIC. Shown as dots are only
     values with errors smaller than 0.20 on both HR(i) and HR(i+1). Foreground
     stars and candidate are marked as big and small stars, galaxies and candidates 
     as big and small x, AGN candidates as
     crosses, SSS candidates as triangles, SNR and candidates as big and small
     hexagons, XRBs as squares. In addition, we mark positions derived from
     measured \xmm\ EPIC spectra and models  for SSS (S1 to S4) as filled
     triangles, low mass XRBs (L1 and L2) as filled squares, SNRs (N132D as N1,
     1E~0102.2-7219 as N2, N157B as N3, Crab spectra as C1 and C2) as filled hexagons, 
     AGN (A1 and A2) as asterix.
     }
    \label{hr} 
\end{figure}
\begin{table*}
\begin{center}
\caption[]{Summary of identifications and classifications.}
\begin{tabular}{llrr}
\hline\noalign{\smallskip}
\hline\noalign{\smallskip}
\multicolumn{1}{c}{Source type} & 
\multicolumn{1}{c}{Selection criteria} &
\multicolumn{1}{c}{identified} &
\multicolumn{1}{c}{classified}  \\ 
\noalign{\smallskip}\hline\noalign{\smallskip}
fg Star & ${\rm log}({{\rm f}_{\rm x} \over {\rm f}_{\rm opt}}) < -1.0$ and HR2 $< 0.3$ and HR3 $< -0.4$ or not defined & 5 & 30 \\
AGN  &  Radio source and not classification as SNR from HR2 &  & 12 \\
GAL  &  optical id with galaxy and HR2 $< 0.0$ & 1 & 1 \\
SSS  &  HR1 $< -0.2$, HR2 - EHR2 $< -0.99$ or HR2 not defined, HR3, HR4 not defined &   &  5 \\
SNR  &  HR1 $> 0.1$ and HR2 $<-0.4$ and not a fg Star & 21 & 23 \\
XRB  &  optical identification or X-ray variability  & 2 &  \\
hard &  HR2-EHR2 $> -0.2$ or only HR3 and\/ HR4 defined, and no other classification&  & 267 \\
\noalign{\smallskip}
\hline
\noalign{\smallskip}
\end{tabular}
\label{class}
\end{center}
\end{table*}
To identify the \m33\ X-ray sources we searched for correlations around the X-ray source positions 
(within a radius of $3\times(\sigma_{stat} + \sigma_{syst})$ in the SIMBAD and NED archives
and within several catalogues. 
In cols. 14 to 19 of Table~\ref{master}, we give extraction information from the USNO-B1 
catalogue (name, number of objects within search area, distance, B2, R2 and I2 magnitude of 
the brightest object). We omitted information, when the digitization was fooled by extended
emission within \HII\ regions (mostly only some colours available and rather bright). 
To improve on the reliability of identifications we used the B and R magnitudes to
calculate the ratio of X-ray-to-optical flux, given by 
${\rm log}({{\rm f}_{\rm x} \over {\rm f}_{\rm opt}}) = {\rm log}({\rm f}_{\rm x}) + ({\rm m}_{\rm B2} + {\rm m}_{\rm R2})/(2\times2.5) + 5.37$, 
following \citet[][ see col. 20]{1988ApJ...326..680M}.

The catalogued X-ray sources are identified 
or classified based on properties in X-rays 
(hardness ratios (HR), variability, extent) and of correlated objects in other 
wavelength regimes (Table~\ref{master}, cols. 21, 22). 
The criteria used are summarized in Table~\ref{class}. We counted a source as identified,
if at least two criteria secure the identification. Otherwise, we only counted a source as 
classified (indicated by pointed brackets). 

We plotted X-ray colour/colour diagrams based on the HRs (see Fig.~\ref{hr}).
Unclassified sources are only plotted if the error in both contributing HRs 
is below 0.2. To identify areas of specific source classes in the plots, we over-plotted 
colours of sources, derived from measured \xmm\ spectra and model simulations (see discussion
in subsections below).

Identification and classification criteria and results are discussed in detail in 
the subsections on individual source classes below. Many foreground stars, SSS and SNRs
can be classified or identified. However, besides a few clearly identified XRBs and AGN,
and SNR candidates with known positions at other wavelengths,
we have no clear hardness ratio criteria (see below) to select XRBs, Crab-like SNR or 
AGN. They are all ``hard" sources and we therefore introduced a class $<$hard$>$ for 
sources with HR2 minus HR2-error greater than -0.2 (see Tab.~\ref{class}). 

43 sources remained unidentified or without classification.

\subsection{Foreground stars}  
Foreground stars are a class of X-ray sources that is homogeneously distributed
over the FOV of the \m33\ survey. The good positioning of \xmm\ and the 
available catalogue USNO-B1 allowed us to effectively select this type of 
sources. We identified five sources with stars of known type in the SIMBAD data
base. Their X-ray fluxes were in the range expected for these sources. In
addition, we classified 30 sources as foreground stars
(${\rm log}({{\rm f}_{\rm x} \over {\rm f}_{\rm opt}}) < -1$, see 
\citet{1988ApJ...326..680M} and in addition HR2 $ < 0.3$ and HR3 $ < -0.4$).
For several of the star candidates we estimated the type from the colours in the
USNO-B1 catalogue using the stellar spectral flux library from  
\citet{1998PASP..110..863P}. For ten candidates we could not determine the type
from the colours (28, 54, 77, 122, 128, 262, 308, 337, 371, 393). This may indicate that
these sources are not isolated stars but more complicated systems or even,
in some cases, galaxies. This expectation could, for instance, be confirmed for
source 262, which was resolved into two sources in DSS2 images.

A special case was source 196, only $\sim2\arcmin\ $ WSW of the nucleus. The source
correlated with 2MASS J01334186+3038491, and was catalogued 
by \citet{1993ApJS...89...85I} as red star 184. The X-ray HRs, optical colour and 
${\rm log}({{\rm f}_{\rm x} \over {\rm f}_{\rm opt}})$ point towards
a foreground star of type M6 as the identification of this source.

From the HRs, sources 7 and 10 should be foreground
stars. However, there was no candidate star within the error box. As these
sources were at the border of the field, the systematic error may be
underestimated and nearby stars may be candidates for the optical 
identification. 

For source 240, the HRs indicated a foreground star, however 
${\rm log}({{\rm f}_{\rm x} \over {\rm f}_{\rm opt}}) = -0.7$, larger than
the assumed limit for classification. The optical colours would indicate a M10 III
star. Also the other candidate stars of type late M have a rather high  
${\rm log}({{\rm f}_{\rm x} \over {\rm f}_{\rm opt}})$.

For several sources,  ${\rm log}({{\rm f}_{\rm x} \over {\rm f}_{\rm opt}})$ 
pointed towards a stellar identification (8, 38, 72, 176, 206, 209, 226, 234). 
However, only for three sources the HRs including errors were within the 
limits we assume for foreground stars (72, 176, 234). 

\subsection{Galaxies and AGN}  
Already after the \ein\ observations, the X-ray bright source X--9c (297) 
was identified by \citet{1982ApJ...253L..13C} with an elliptical galaxy at 
z=0.03. The \xmm\ X-ray luminosity of 1.35\ergs{41} (0.2--4.5 keV) was typical  
and the HRs were compatible with an elliptical galaxy 
spectrum (see HR plots). 
 
From SIMBAD and NED searches, we found an additional galaxy correlation 
(nameley source 358 = LEDA5899). 

There were no correlations with AGN with known redshift within SIMBAD or NED. 
We therefore only classifed
12 sources as AGN based on SIMBAD, NED, and other radio source correlations
(NVSS, \citet{1999ApJS..120..247G}) with the additional condition of being a 
$<$hard$>$ X-ray source. A final decision will only be possible using 
optical spectra. In Fig.~\ref{hr}, we included typical HRs for an AGN spectrum
(power law with photon index of 1.7 assuming galactic foreground absorption
(6\hcm{20}, A1) and absorption through the densest areas of \m33\ 
\citep[][ 4.8\hcm{21}, A2]{1980MNRAS.191..615N}.

Sources 150 and 373 could be classified as SNRs, as well as AGN. They were identified as 
radio sources within the inclination-corrected \m33\ D$_{25}$ ellipse, without optical 
identification as SNR by \citet{1999ApJS..120..247G} and
the radio spectral index did not clearly classify them as AGN.

\subsection{Super-Soft Sources (SSS)}
SSS show black body spectra with temperatures below 50 eV, they radiate close to the
Eddington luminosity of a 1 M\sun\ object and they are believed to be white dwarf
systems steadily burning hydrogen at the surface. They were identified as a
class of X-ray sources by ROSAT and are often observed as
transient X-ray sources 
\citep[see][ and references therein]{2000NewA....5..137G}.
In the catalogue, SSS were only classified using their HRs. 
To guide the classification, we calculated SSS HRs assuming a 25 
and 50 keV black body model spectrum assuming galactic foreground 
absorption (S1 and S3) and absorption through the densest areas of \m33\ 
(S2 and S4).  This led to the selection criteria in Table~\ref{class} and
the classification of five \m33\ SSS (52, 109, 223, 247, 332). They were detected 
with absorbed luminosities of (1--8)\ergs{35} in the 0.2-0.5 keV band. Only
three of the sources (and the model HRs) could be plotted in Fig.~\ref{hr}. 
For the other two, HR2 was not determined. 

\subsection{Supernova remnants (SNR)}
SNRs can be identified as sources where thermal components
dominate the X-ray spectrum below 2 keV (examples are N132D in the Large
Magellanic Cloud and 1E~0102.2-7219 in the Small Magellanic Cloud) and as
so called ``plerions" with power law spectra (examples are the Crab nebula and
N157B in the Large Magellanic Cloud). To guide the classification we calculated
HRs from archival \xmm\ spectra of these SNRs. Spectra of N132D (N1), 
1E~0102.2-7219 (N2), and N157B (N3) 
could be directly compared to \m33\ SNRs, as they are seen through 
comparable foreground absorption. For the Crab nebula spectrum, we assumed 
galactic foreground absorption (C1) and absorption through the densest areas 
of \m33\ (C2). It is clear from Fig.~\ref{hr}, that ``thermal" SNRs are located
in areas of the X-ray colour/colour plots that only overlap with foreground stars. If we
assumed that we have identified all foreground star candidates from the optical
correlation and inspection of the optical images, the remaining sources could be
classified as SNRs with the criteria given in  Table~\ref{class}. 

Extensive searches for SNRs in \m33\ were performed in the optical 
\citep[see][ and references therein]{1998ApJS..117...89G}
and radio
\citep[see][ and references therein]{1999ApJS..120..247G}. Many X-ray
sources in the catalogue correlated with optical and/or radio SNRs (see
Tab.~\ref{master}). We counted such sources as SNR identifications (21), if 
in addition they fulfilled the HR criteria (see above). We counted sources as
classified SNRs (23), if they either just
fulfilled the HR criteria (16) and were no identified or classified as foreground star,  
or if they correlated with optical/radio SNRs and did not fulfill the HR criteria (7).
 
From these classified SNRs, sources 20 fulfilled the HR criteria,  
but was positioned outside the inclination-corrected optical D$_{25}$ ellipse,
while sources 101 and 266 were located between two  optical SNR candidates 
and may represent the combined emission of both.
As already mentioned in the subsection on AGN, we classify sources 150 and 373
as SNR as well as AGN due to the radio correlation.

A special case was source 227, which correlated with radio source 126 of 
\citet{1999ApJS..120..247G}, who classified it as a background source with
no optical identification. The \xmm\ HRs clearly indicated a SNR nature and we
therefore classified the source as a SNR.

The giant \HII\ regions BCLMP 290 and NGC~604 were detected as extended X-ray 
sources (125 and 299). While the HRs of source 299 indicated a spectrum
similar to thermal SNRs (several SNRs would be necessary to explain the
X-ray emission), the HRs of source 125 indicated a hard spectrum which may
originate from several point sources (XRBs) or Crab-like SNRs. In the
catalogue we only classified source 299 as a SNR.

Four additional sources in the catalogue (137, 154, 162, 280) had HRs 
compatible with the selection criteria in Table~\ref{class}, within the
errors. They are good candidates for further SNR investigations in \m33.

\subsection{X-ray binaries (XRB)}
As already mentioned in the introduction to this section, 
expected spectra of XRBs are similar to AGN and
Crab-like SNRs. To guide the classification, we calculated
HRs as expected for low mass XRBs (5 keV thermal Bremsstrahlung spectrum 
assuming galactic foreground 
absorption (L1) and absorption through the densest areas of \m33\ (L2)), which
show soft spectra for XRBs. 
As can be seen in Fig.~\ref{hr} these different source classes did not separate. 

Therefore we only listed the black hole XRB  X--8  close to the nucleus (211) 
and the 3.4~d eclipsing HMXB X--7 (171) as identified X-ray binaries.

SIMBAD searches indicated 16 more good XRB candidates from positional
coincidences with 
optical stars of different type in \m33\ (see Table~\ref{master}). Source 110
correlated with an emission line object and is a good candidate for a Be
XRB, source 174 correlated with an eclipsing binary candidate (398 d period),
and sources 194, 197, and 323  with variable stars. Of historical
interest is the optical candidate for source 323, VHK6. This source was
discussed as one of four irregular variables in the paper by
\citet{1926ApJ....63..236H}, in which he proposed the spiral nebula \m33\ as a 
stellar system outside the Milky Way.   

All the candidates mentioned above need further X-ray work to confirm their
XRB nature.
\section{Discussion}
\begin{figure}
    \caption[]{\xmm\ EPIC 0.2-2 keV band contours of 
    (6,8,14,32,86)$\times 10^{-5}$ ct s$^{-1}$
    overlayed on an optical image of \m33\ extracted from the digitized sky survey (DSS2) red
    plates. The image scale is marked. 
    \label{opt}}
\end{figure}
Besides the large number of point-like sources, there was a strong component of
unresolved X-ray emission in the energy band below 2 keV. The overlay of 
X-ray on the DSS2 red image (Fig.~\ref{opt}) clearly showed the strong
correlation with the inner disk and the southern and NW spiral arm. While some
of this emission certainly is due to unresolved point sources, there 
is a significant contribution from hot gas in the disk which is more 
concentrated in the spiral arms and \HII\ regions like NGC~604 12\arcmin\ NE and 
BCLMP 290 15\arcmin\ NW of the nucleus. A detailed analysis of
this emission will be presented in a separate paper.  

In the \xmm\ source catalogue we presented 408 sources in the \m33\ field, 35 of
which could be identified or classified as foreground stars and 14 as background AGN or
galaxies. As sources in \m33\ we identified or classified 5 SSS, 44 SNRs, and 2
XRBs. 267 sources were classified as hard and may be either XRBs or Crab-like
SNRs in \m33\ or background AGN. There remained 43 sources unclassified.

We can estimate the contribution of background sources to the source catalogue
extrapolating from X-ray deep field number counts in the 0.5--2.0 keV and 2.0--8.0 keV bands
\citep[see e.g.][]{2001AJ....122.2810B,2002ApJ...566..667R}. We used 
rough sensitivity maps for our survey created by the SAS task {\tt
esensmap} for the 0.5--2.0 keV and 2.0--4.5 keV bands
and an estimate for additional absorption in the line-of-sight. 
One expects about 175 to 210 background sources detected in our field-of-view,
if we assume an overlap of sources detected in the hard and soft band of 75\% or 
only 50\%. Fourteen of the background sources have already been identified 
(see Sect. 6.2). Taking these numbers into account, up to 50\% of the
hard or unclassified sources should be sources within \m33\ (mostly XRBs or Crab-like SNRs).

We compared the \xmm\ source catalogue with the ROSAT catalogue of HP01.  
While most of the bright ROSAT sources were also detected by \xmm\, many of 
the sources, only detected with the ROSAT HRI with a detection
likelihood below 10, were not detected by \xmm\ and seem to be spurious 
detections. From the seven ROSAT sources classified as SSS, we did not detect
four. As expected for this source class, these might be transients, that were not active
during the \xmm\ observations (HP01 sources 43, 69, 76, 82). The remaining three 
seem to be false identifications: HP01 source 61 is now classified by \xmm\ as a SNR, 
84 as a foreground star and 124 as a hard source. Two ROSAT correlations with radio
sources are no longer valid with the improved \xmm\ positions (HP01 sources 63 and
147). 

HP01 identified and classified 17 SNRs. The \xmm\ source catalogue
marks 21 identifications and 23 classifications. While the HR criteria seem to
be rather robust for thermal SNRs, the identification of Crab-like remnants is
doubtful and can only be secured by measurement of source extent. This may be
possible with \xmm\ EPIC for very extended SNRs (optical extent of SNR in \m33\ 
vary from 9\arcsec\ to 126\arcsec\ according to \citet{1998ApJS..117...89G}). Most
of the sources, however, will need dedicated \chandra\ on-axis observations for a final
decision on the extent. The \chandra\ count rates will not be sufficient to get
good spectra and a combined \xmm\ and \chandra\ analysis will be important
for this class of sources.

The importance of combined  \xmm\ and \chandra\ data for \m33\ XRB research
was demonstrated in \citet{px7}. This type of analysis
extended to all brighter sources will be  the subject of a future paper.

\begin{acknowledgements}
This publication makes use of the USNOFS Image and Catalogue Archive
operated by the United States Naval Observatory, Flagstaff Station
(http://www.nofs.navy.mil/data/fchpix/), 
of data products from the Two Micron All Sky Survey, 
which is a joint project of the University of Massachusetts and the Infrared 
Processing and Analysis Center/California Institute of Technology, funded by 
the National Aeronautics and Space Administration and the National Science 
Foundation, of the SIMBAD database,
operated at CDS, Strasbourg, France, 
and of the NASA/IPAC Extragalactic Database (NED) 
which is operated by the Jet Propulsion Laboratory, California 
Institute of Technology, under contract
with the National Aeronautics and Space Administration.
The compressed files of the ``Palomar Observatory - Space Telescope Science Institute 
Digital Sky Survey'' of the northern sky, based on scans of the Second Palomar Sky 
Survey, are copyright (c) 1993-1995 by the California Institute of
Technology and are distributed herein by agreement. All Rights Reserved.
Produced under Contract No. NAS 5-26555 with the National Aeronautics and Space 
Administration. 
The \xmm\ project is supported by the Bundesministerium f\"{u}r
Bildung und Forschung / Deutsches Zentrum f\"{u}r Luft- und Raumfahrt 
(BMBF/DLR), the Max-Planck Society and the Heidenhain-Stiftung.
\end{acknowledgements}

\bibliographystyle{aa}
\bibliography{./0068,/home/wnp/data1/papers/my1990,/home/wnp/data1/papers/my2000,/home/wnp/data1/papers/my2001}

\end{document}